\documentclass{article}
\usepackage{PRIMEarxiv}
\usepackage{authblk}
\usepackage[utf8]{inputenc} 
\usepackage[T1]{fontenc}    
\usepackage{hyperref}       
\usepackage{url}            
\usepackage{booktabs}       
\usepackage{amsfonts}       
\usepackage{nicefrac}       
\usepackage{microtype}      
\usepackage{lipsum}
\usepackage{fancyhdr}       
\usepackage{graphicx}       
\usepackage[backend=biber, style=numeric, maxbibnames=3, mincitenames=1, sorting=none, labelnumber, bibstyle=numeric]{biblatex}

\title{Three-Dimensional Amyloid-Beta PET Synthesis from Structural MRI with Conditional Generative Adversarial Networks
\thanks{2024 International Society of Magnetic Resonance in Medicine.  Singapore, Singapore,  May 4-9.  Abstract Number 2239}}
\author[1,2,4]{Fernando Vega}
\author[1,2,4]{Abdoljalil Addeh}
\author[1,2,3,4]{M. Ethan MacDonald}

\affil[1]{Department of Biomedical Engineering, Schulich School of Engineering, University of Calgary, AB, Canada}
\affil[2]{Department of Electrical \& Software Engineering, University of Calgary, Calgary, AB, Canada}
\affil[3]{Department of Radiology, Cumming School of Medicine, University of Calgary, Calgary, AB, Canada}
\affil[4]{Hotchkiss Brain Institute, University of Calgary, Calgary, AB, Canada}

\date{May 2024}

\begin{document}

\maketitle

\section*{Synopsis}
\textbf{Motivation:} Alzheimer’s Disease hallmarks include amyloid-beta deposits and brain atrophy, detectable via PET and MRI scans, respectively.  PET is expensive, invasive and exposes patients to ionizing radiation.  MRI is cheaper, non-invasive, and free from ionizing radiation but limited to measuring brain atrophy.

\noindent \textbf{Goal:} To develop an 3D image translation model that synthesizes amyloid-beta PET images from T1-weighted MRI, exploiting the known relationship between amyloid-beta and brain atrophy.

\noindent \textbf{Approach:} The model was trained on 616 PET/MRI pairs and validated with 264 pairs.

\noindent \textbf{Results:} The model synthesized amyloid-beta PET images from T1-weighted MRI with high-degree of similarity showing high SSIM and PSNR metrics (SSIM>0.95\&PSNR>28).

\noindent \textbf{Impact:} Our model proves the feasibility of synthesizing amyloid-beta PET images from structural MRI ones, significantly enhancing accessibility for large-cohort studies and early dementia detection, while also reducing cost, invasiveness, and radiation exposure.

\section*{Introduction}

Alzheimer’s Disease (AD) is the predominant form of dementia \cites{AD}.  AD is a progressive neurodegenerative disease characterized by memory impairment and cognitive decline \cites{kumar2022}.  Dementia had an enormous global economic burden of \$1.3 trillion USD in 2019 \cite{econo}.  Histopathologically, AD is characterized by the deposition of amyloid-beta within the brain, a key molecule associated with AD progression \cite{amyloidBeta}.  When amyloid-beta accumulates between neurons, it can lead to synaptic failure and neuronal death, manifesting as brain atrophy \cite{Inflammation}. 

In recent decades,  imaging amyloid-beta has become possible using positron emission tomography (PET) with radiotracers such as Pittsburgh Compound B (PiB) \cite{Yamin2017}.  However, PET imaging is costly (\$5000-\$8000 per scan) \cite{Wittenberg2019}, invasive (requiring injection of a radiotracer), and involves exposure to harmful ionizing radiation \cite{Marti-Climent2017}.  These factors limit its availability in many jurisdictions and make it challenging to use for early-onset AD detection. 
Structural MRI is a more affordable, non-invasive, alternative to PET that does not use ionizing radiation \cite{McMahon2011}.  While it primarily targets brain atrophy driven by AD, MRI cannot provide amyloid-beta measurements as PET does \cite{Vemuri2010}.  Addressing this limitation, image translation models have been developed to generate synthetic amyloid-beta PET images from structural MRI \cite{BPGAN, GLA-GAN}, using the known relationship between amyloid-beta burden and brain atrophy \cite{spattialpatterns, relationab}. 

Image translation models can generate synthetic images from other image types, providing access to images that might be difficult to obtain \cite{mygod}.  These models are commonly implemented with Conditional Generative Adversarial Networks (cGAN) due to their ability to generate realistic synthetic images \cite{imagetrans}.  Prior research has demonstrated success in generating 2D axial amyloid-beta PET images from structural MRI \cite{Vega}.

This work builds on the previous study by introducing a 3D-cGAN model with significant architectural improvements, enhancing the amyloid-beta PET synthesis shown in the image comparison metrics.

\section*{Methods}

The Open Access Series of Imaging Studies (OASIS-3) \cite{OASIS} dataset included 1098 subjects with both PiB PET and MRI images, comprising 609 cognitively normal (CN) individuals and 489 at different levels of cognitive decline.  The MRI images were captured using three scanners: Siemens Biograph mMR 3T, Siemens Trio Tim 3T, and Siemens Sonata 1.5T with a resolution of 1x1x1mm.  The amyloid-beta images were acquired with two scanners: Siemens Biograph 40 PET/CT and Siemens ECAT 962 using PiB radiotracer with an injected dose ranging from 6 to 20 millicuries and a 60-minutes dynamic scan and standard uptake value ration (SUVR) was obtained using the PET Unified Pipeline. 

The MRI images were preprocessed with FreeSurfer to produce brain extractions.  Dynamic PET images were converted to static by summing frames between 30 and 60 minutes.  These images were co-registered to the MRI images using an affine registration with the Advanced Normalization Tools (ANTs), and brain extraction was performed using the MRI as a mask.  PET images were SUVR normalized by dividing the mean tracer concentration in the cerebellum cortex, a standard reference region for PiB PET imaging \cite{kinetic, Jack}.

\begin{figure}[h]
\includegraphics[width=\textwidth]{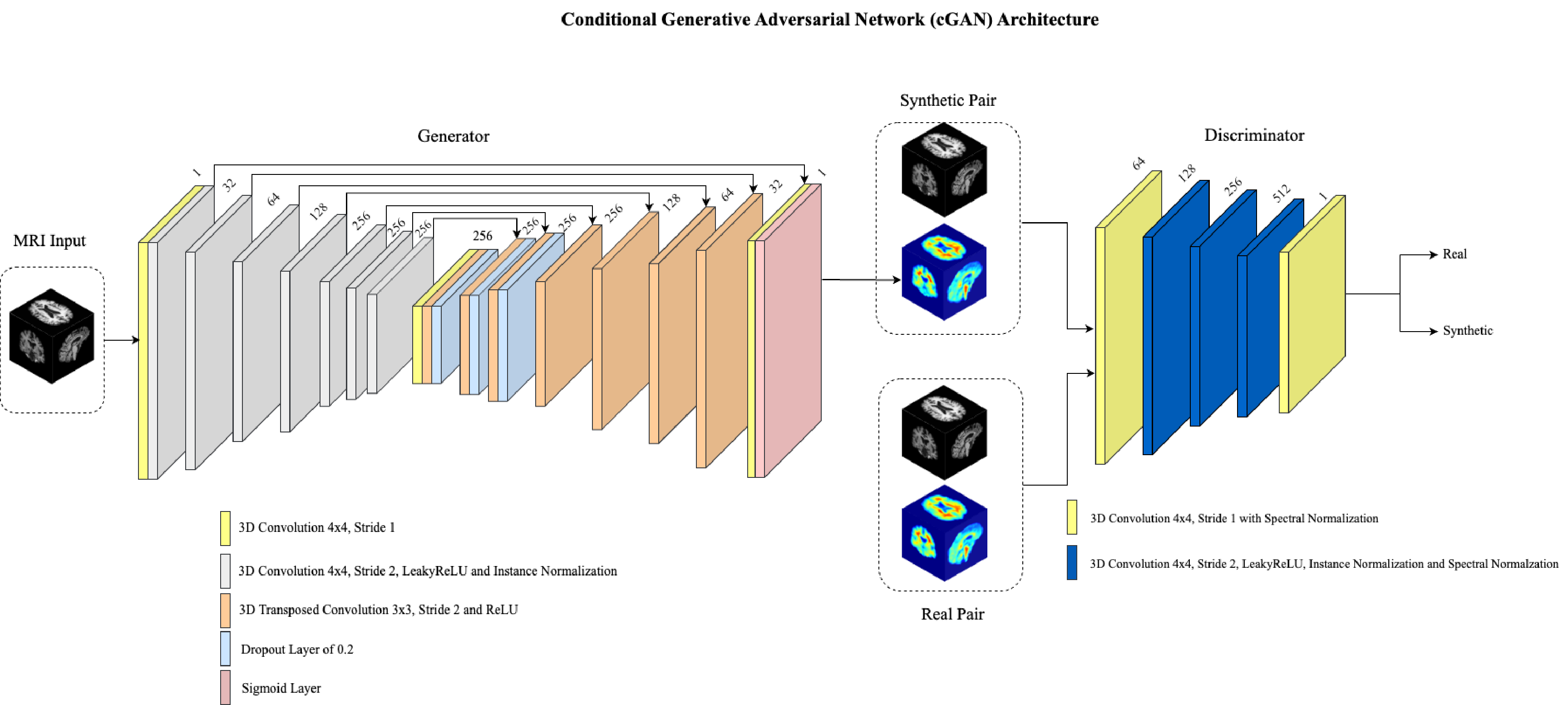}
\caption[+30pt]{Proposed image translation model following a 3D-cGAN architecture. The generator uses an encoder-decoder architecture that receives an MRI input and generates a synthetic PET image (output) that is compared with the real PET (label).  Then the real pair (real PET and MRI input) and synthetic pair (synthetic PET and MRI input) are independently fed into the discriminator which classifies the given pair as real or synthetic. The discriminator incorporated spectral normalization in the middle layers.} 
\label{model_arch}
\end{figure}

The 3D-cGAN architecture as shown in Figure \ref{model_arch} was implemented using MONAI.  A brain area mask loss was incorporated to focus on brain information while disregarding the background.  To prevent exploding gradients, spectral normalization was applied in the discriminator.  PET images were z-score normalized to a range between 0 and 1, with a mean=0 and standard-deviation=49.72.  This normalization facilitates denormalization without compromising tracer quantification and improves the stability of the training process.

The model was trained with 616 PET/MRI pairs and validated with 264 pairs.  The training and validation cohorts were stratified with equal proportion of females/males and different levels of cognitive decline.

The validation cohort was evaluated with Structural Similarity Index Measure (SSIM) and Peak Signal-to-Noise Ratio (PSNR). 

\newpage
\section*{Results}
\begin{figure}[h]
\centering
\includegraphics[width=0.8\textwidth]{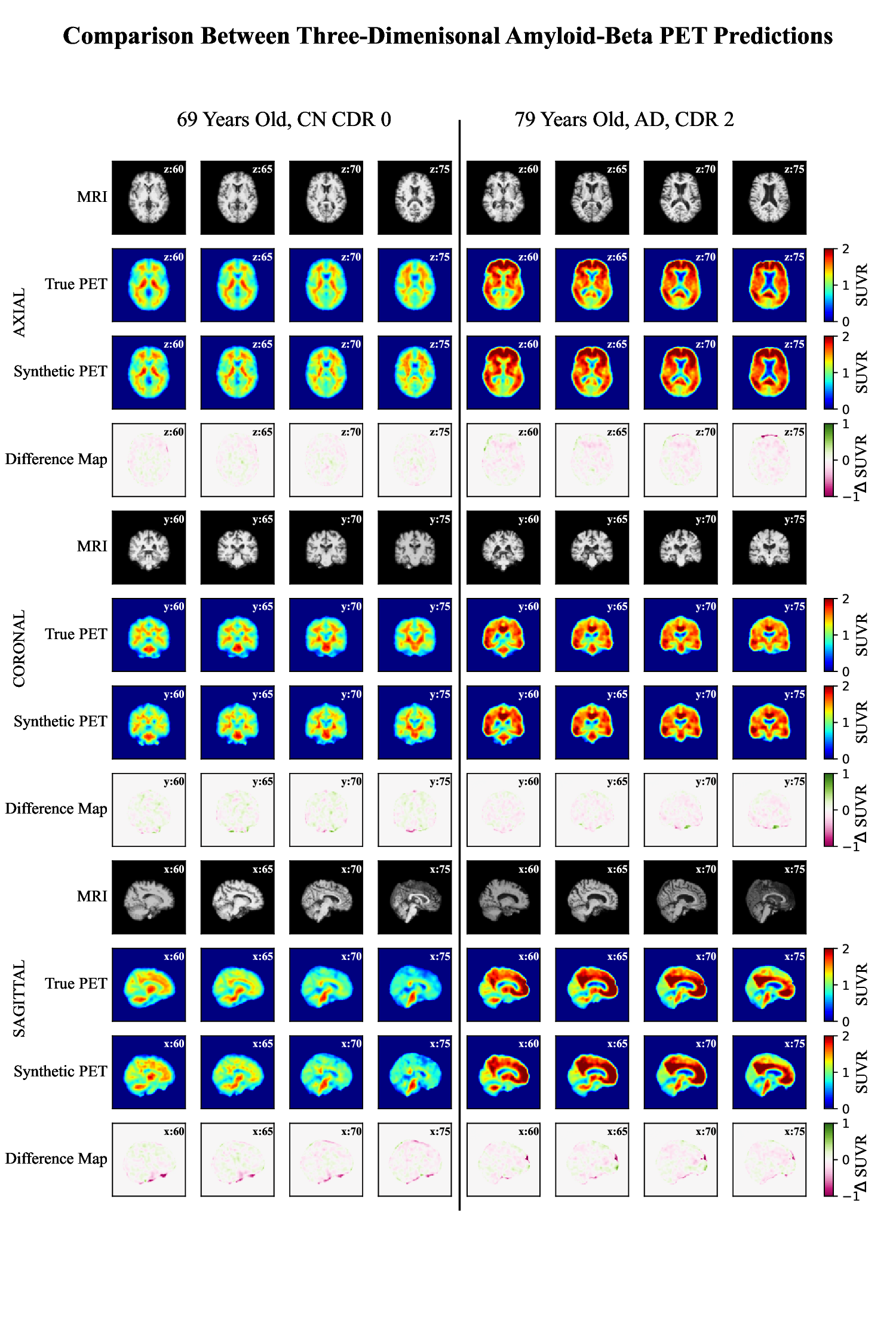}
\vspace{-2cm}
\caption{Amyloid-beta synthesis comparison. Two subjects are compared, a cognitively normal (CN) and an Alzheimer’s Disease (AD) across three anatomical views: axial, coronal and sagittal. For each view an MRI, real PET, synthetic PET and a difference map between real and synthetic PET. Showing that the model can produce high-quality synthetic PET images for CN and AD cases that are close in shape and SUVR quantification.} 
\label{comparison}
\end{figure}

\begin{figure}[p]
\centering
\includegraphics[width=0.8\textwidth]{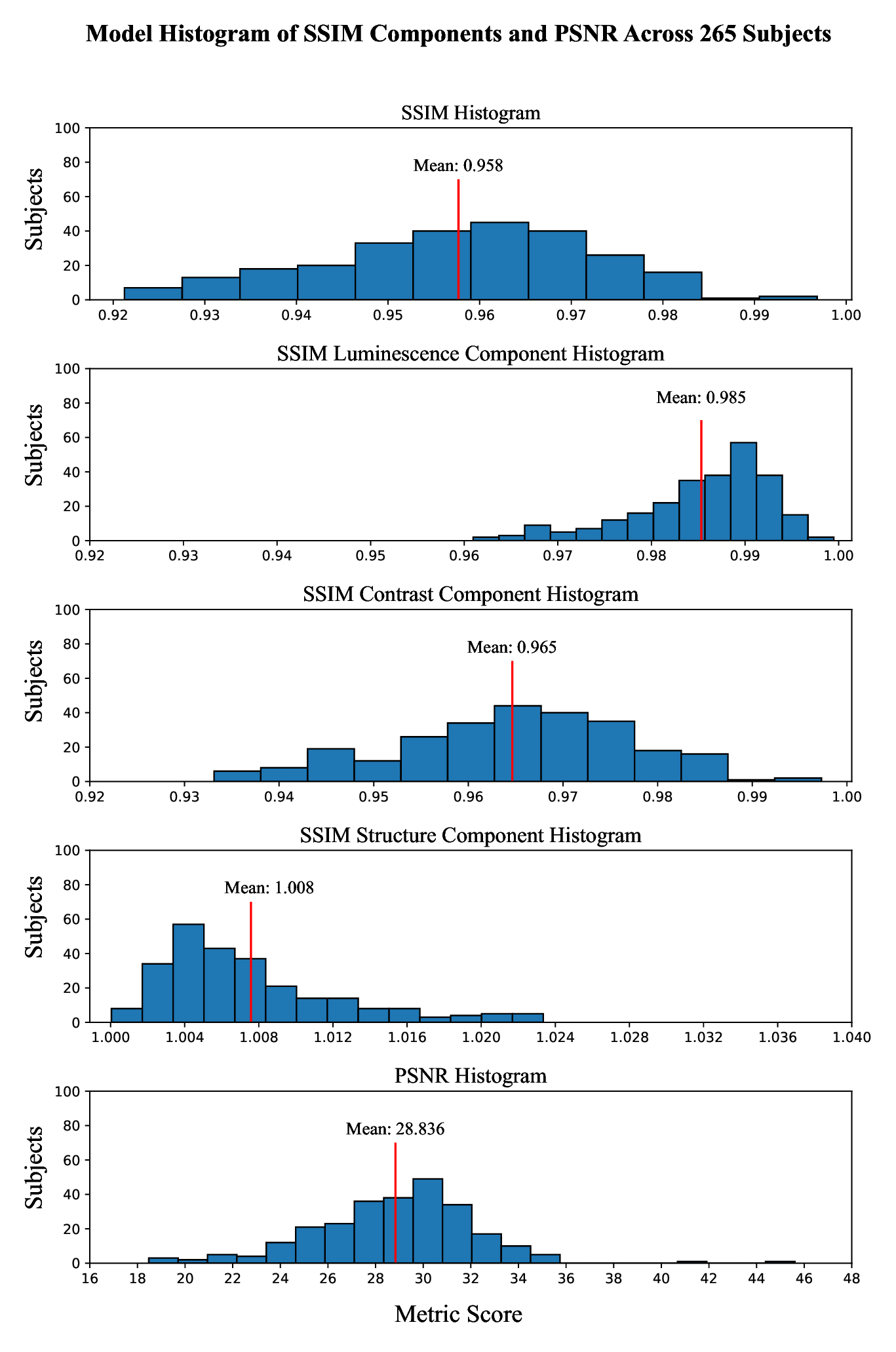}
\vspace{-0.5cm}
\caption{SSIM and PSNR data distribution across 264 cohort of unseen subjects. SSIM and its luminescence, contrast, and structure components are reported, reaching a mean SSIM above 0.95, mean luminescence component above 0.98, mean contrast component above 0.96 and mean structure component above 1. With a mean PSNR above 28 indicating that the model can produce synthetic PET images have high degree of similarity with the true ones.} 
\label{histograms}
\end{figure}

Figure \ref{comparison} showcases two representative subjects, illustrating the model’s ability to synthesize high-fidelity amyloid-beta PiB PET images from T1-weighted MRI for both CN and AD cases, presented in three anatomical planes.  Figure \ref{histograms} displays the distribution of SSIM in contrast, luminescence, and structural components, alongside the PSNR.  The model achieved a mean SSIM=0.958 and PSNR=28.836

\section*{Discussion}
\begin{figure}[h]
\includegraphics[width=\textwidth]{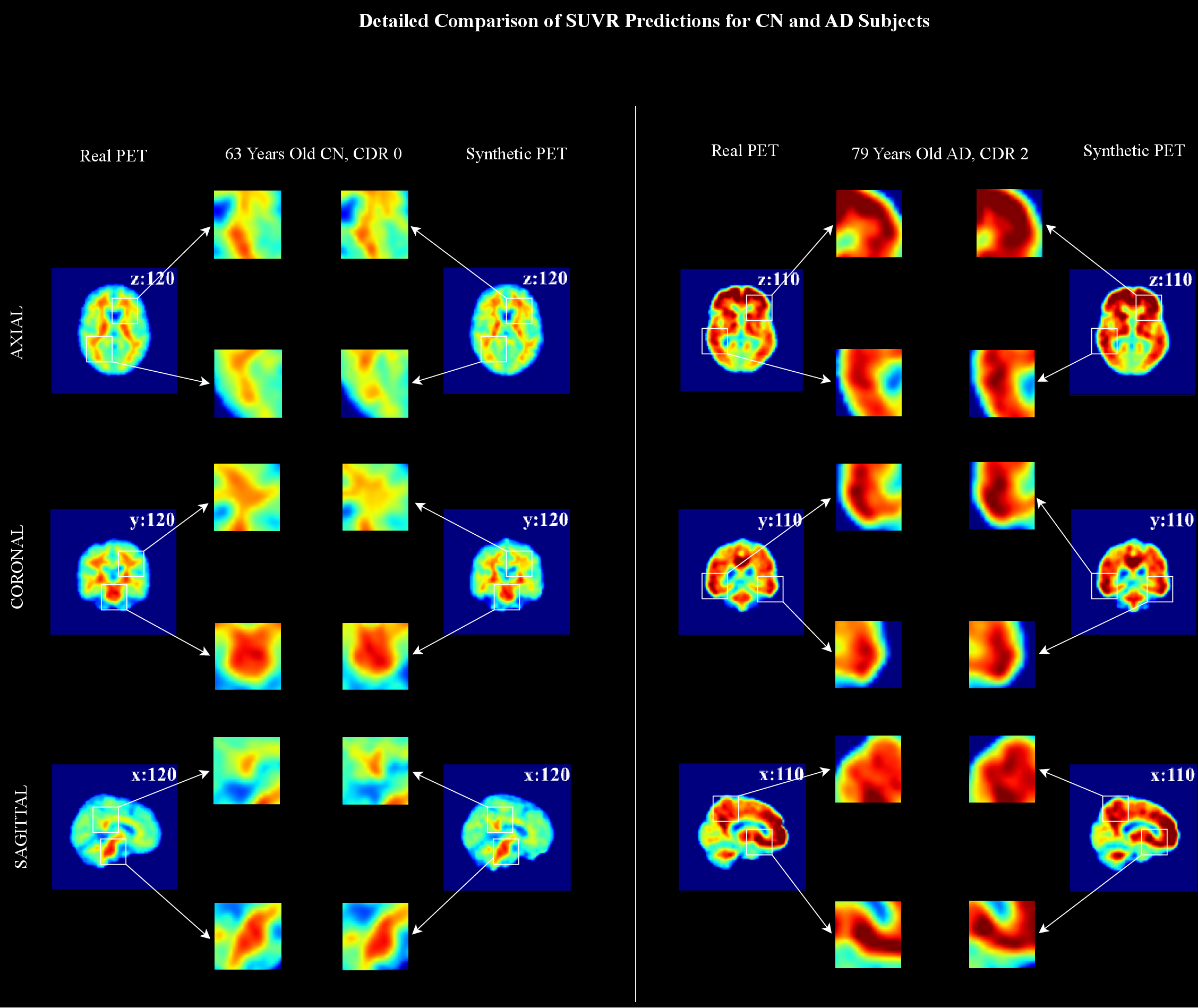}
\centering
\caption{Detailed SUVR comparison between synthetic and real amyloid-beta PET images, showing the axial, coronal and sagittal planes. While the synthetic images highly resemble the real ones, there is still room for improvement. The AD case in the axial plane, shows that the tracer accuracy requires enhancements while the CN subject shows higher accuracy in the SUVR quantification.} 
\label{final_comp}
\end{figure}

Future work will aim to improve the precision of the synthetic PET images, particularly for subtle discrepancies shown in Figure \ref{final_comp}.  This can be addressed by exploring model architectures such as Self-Attention Conditional GAN, vision transformers GAN, and stable diffusion models.  A prevalent challenge in GAN is the emphasis on relative image quality at the expenses of quantitative contrast accuracy.  Future work will include developing metrics that better quantify contrast fidelity. 

\section*{Conclusion}
This study demonstrates that a three-dimensional image translation model is capable of synthesizing amyloid-beta PiB PET images from structural MRI with high-degree of similarity, achieving mean SSIM>0.95 and PSNR>28.  This approach offers a cost-effective, less-invasive, and radiation-free method for early AD screening.  

\newpage
\section*{Acknowledgments}
The authors would like to thank the University of Calgary, in particular the Schulich School of Engineering and Departments of Biomedical Engineering and Electrical \& Software Engineering; the Cumming School of Medicine and the Departments of Radiology and Clinical Neurosciences; as well as the Hotchkiss Brain Institute, Research Computing Services and the Digital Alliance of Canada for providing resources.  The authors would like to thank the Open Access of Imaging Studies Team for making the data available.  JA – is funded in part from a graduate scholarship from the Natural Sciences and Engineering Research Council Brain Create.  MEM acknowledges support from Start-up funding at UCalgary and a Natural Sciences and Engineering Research Council Discovery Grant (RGPIN-03552) and Early Career Researcher Supplement (DGECR-00124), and funding from the Natural Sciences and Engineering Research Council Alliance Advance and Alberta Innovates.  This work was made possible through a generous donation by Jim Gwynne. 

\printbibliography

\end{document}